\documentstyle[11pt,fleqn]{article}
\newcommand{\nn}{\nonumber}            
\renewcommand{\Re}{\,\mbox{Re}\,}                

\newcommand{\de}{\delta}
\newcommand{\ep}{\varepsilon}

\begin{document}
\parindent0mm

\renewcommand{\theequation}{\mbox{\arabic{section}.\arabic{equation}}}
\newcommand{\reals}{\mbox{${\rm I\!R }$}}
\newcommand{\nats}{\mbox{${\rm I\!N }$}}
\newcommand{\intgs}{\mbox{${\rm Z\!\!Z }$}}
\newcommand{\komplex}{\mbox{${\rm I\!\!\!C }$}}
\newcommand{\beq}{\begin{eqnarray}}
\newcommand{\eeq}{\end{eqnarray}}
\newcommand{\sumh}{\sum_{\{P\}_{\Gamma}}}
\newcommand{\sume}{\sum_{\{Q\}_{\Gamma}}}
\newcommand{\rd}{(r^2+\delta^2)}
\newcommand{\erd}{\left(1+\frac{r^2}{\delta^2}\right)}
\newcommand{\sik}{\sin\left(\frac{k\pi} n\right)}
\newcommand{\iou}{\int_0^{\infty}}
\newcommand{\iuu}{\int_{-\infty}^{\infty}}
\newcommand{\siqk}{\sin^2\left(\frac{k\pi} n\right)}
\newcommand{\ch}{\chi^k(P)}
\newcommand{\chq}{\chi^k(Q)}
\newcommand{\sumkeu}{\sum_{k=1}^{\infty}}
\newcommand{\sumkem}{\sum_{k=1}^{n-1}}
\newcommand{\abl}{\partial}
\newcommand{\lm}{\ln\left(\frac{\lambda M^2}{2\mu ^2}\right)}
\newcommand{\slnuun}{\sum_{l_1,...,l_N=-\infty}^{\infty}
                 \!\!\!\!\!\!\!\!^{\prime}}

\title{Self-interacting scalar fields on spacetime with compact
hyperbolic spatial part}
\author{Andrei Bytsenko\thanks{Permanent address:
Department of Theoretical Physics,
State Technical University, St.~Petersburg 195251, Russia},
Klaus Kirsten\\
{\it Dipartimento di Fisica, Universit\`a degli Studi di Trento}\\
{\it 38050 Povo (Trento), Italia}\\
$$\\
Sergei Odintsov\thanks{On leave from Tomsk Pedagogical Institute,
634041 Tomsk, Russia}\\
{\it Department of Physics, Faculty of Science, Hiroshima
University}\\
{\it Higashi-Hiroshima 724, Japan}}
\date{March 1993}
\maketitle
\begin{tabbing}
Subject classification number:\quad\= 1100\\
                                     \>0460\\
				     \>0230
\end{tabbing}
\begin{abstract}
We calculate the one-loop effective potential of a self-interacting
scalar field on the
spacetime of the form $\reals^2\times H^2/\Gamma$. The Selberg
trace formula associated with a co-compact discrete group $\Gamma$ in
$PSL(2,\reals )$ (hyperbolic and elliptic elements only) is used.
The closed form for the
one-loop unrenormalized and renormalized effective potentials
is given. The influence of non-trivial topology on curvature induced
phase transitions is also discussed.
\end{abstract}
\section{Introduction}
The effective action in quantum field theory in curved spacetime is
the central object which, in principle, should define the dynamics of
the early universe at scales bigger than the Planck scale. In
particular, it should be relevant for the description of the
inflationary universe (for a review see [1-4])
\nocite{kolbturner90}
\nocite{kolb91} \nocite{linde90} \nocite{olive90}
which may be based on
some phase transition (for example induced by temperature [1-4]).
\nocite{kolbturner90}
\nocite{kolb91} \nocite{linde90} \nocite{olive90}
Unfortunately
there are a lot of difficulties
in calculating the effective action in a quantum field theory
in a general curved spacetime (for a review see
\cite{buchbinderodintsovshapiro92}). Therefore it is very natural
to deal with some specific spaces which are interesting from the
cosmological viewpoint.
In this context it is significant to investigate the constant
curvature compact spaces, contained in the well known class of the
so-called Robertson-Walker space-time forms.

The evaluation of the
effective potential
in self-interacting scalar field theories
on curved backgrounds has a long history and
many explicit calculations of effective potentials in different
spacetimes have been done [6-20]
\nocite{denardospallucci80}
\nocite{oconnorhu84} \nocite{oconnorhu86} \nocite{oconnorhushen83}
\nocite{futamase84} \nocite{huang90} \nocite{berkin92}
\nocite{kennedy81,actor90a,elizalderomeo90a,fordyoshimura79,ford80}
\nocite{toms80,hosotani84,shore80a}
(for a review see \cite{buchbinderodintsovshapiro92}).
The influence of nontrivial topology on the effective potential has
been particularly studied in
\cite{ford80,toms80,hosotani84,denardospallucci80,actor90a,elizalderomeo90a},
concentrating however on spacetimes with vanishing curvature. Our
main aim is, to analyze the influence of nonvanishing constant
curvature together with nontrivial topology (see also
\cite{kennedy81,changdowker92unv}). As an example we choose the
spacetime $\reals ^2 \times H^2/\Gamma$, where $\Gamma$ is a
co-compact discrete group in $PSL(2,\reals )$ with hyperbolic and
elliptic elements. For this spacetime the Selberg trace formula
associated with $\Gamma$ allows us to analyze several properties of
the one-loop effective potential.
(Note that recently the classification of topologies of hyperbolic
universes has been discussed in ref.~\cite{koike}.)

The organisation of the paper is as follows. In section 2 the
non-renormalized one-loop effective potential for a quartic
self-interacting scalar field coupled to gravity is calculated using
the Selberg trace formula. In section 3 the renormalization of the
one-loop effective potential is performed. The analysis of the phase
transition in the model is given in section 4, including some remarks
on scalar electrodynamics on $\reals ^2 \times H^2/\Gamma$. Finally a
short summary of our results is given in section 5.
\section{Regularized one-loop effective potential}
\setcounter{equation}{0}
The concept of the effective action and the effective potential is
well discussed in the literature [23-26],
\nocite{brandenberger85}
\nocite{buchbinderodintsovshapiro89} \nocite{colemanweinberg73}
\nocite{jackiw74}
so the introduction of these quantities will be very
brief. We consider a self-interacting scalar field coupled to gravity
on the spacetime of the form
${\cal M}=\reals ^2 \times H^2/\Gamma$.
Here the group $\Gamma$ in $PSL(2,\reals )=SL(2,\reals )/\{ -1,1\}$ is
a discrete co-compact group acting on the two-dimensional Lobachevsky
space $H^2$, i.e.~the signature of $\Gamma$ contains only hyperbolic
and elliptic numbers.
The bare
Lagrangian density of the theory reads
\beq
L(x)=-\frac 1 2 (\nabla_{\mu}\phi)(\nabla^{\mu}\phi)-
\frac 1 2 (m^2+\xi
R)\phi^2-\frac{\lambda }{4!} \phi ^4\label{1}
\eeq
where $m$ is the mass of the field, $\xi$ is the conformal coupling
constant of the quantum
scalar field $\phi$ to the gravitational field, represented
by the scalar curvature $R$. Because $H^2/\Gamma$ is a constant
curvature space, the concept of the effective potential is well
defined \cite{berkin92}. Formally it is given in the form
\beq
V(\phi_c)=V^{(0)}(\phi_c)+\hbar V^{(1)}(\phi_c)+{\cal O}(\hbar
^2)\label{2}
\eeq
with the purely classical part
\beq
V^{(0)}(\phi_c)=\frac 1 2 m^2\phi_c^2+\frac 1 2 \xi R
\phi^2_c+\frac{\lambda}{4!}\phi_c^4\label{3}
\eeq
and with the one-loop quantum corrections of the form
\beq
V^{(1)}(\phi_c)=\frac 1 {2 vol({\cal M})}\ln\det(\mu^2A)\label{4}
\eeq
where $\mu$ is a normalization constant with dimension of length and
the relevant Laplace-like operator $A$ is given by
\beq
A=-\Delta +m^2+\xi R +\frac{\lambda} 2 \phi_c^2\label{5}
\eeq
Using the zeta function prescription for the regularization of
functional determinants \cite{hawking77}, \cite{critchleydowker76},
equation (\ref{4}) is given in the form (from now on we will write
$\phi$ instead of $\phi_c$)
\beq
V^{(1)}(\phi) =-\frac 1
{2vol({\cal M})}\left[{\zeta'}_A(0)+\zeta_A(0)\ln\mu^2\right]\label{6}
\eeq
where $\zeta_A(s)$ is the zeta function associated with the operator
$A$. This means
\beq
\zeta_A(s)=\sum_{\vec k} \Omega_{\vec k}^{-s}\label{7}
\eeq
with the eigenvalues $\Omega_{\vec k}$, $\vec k =(j,n,l)$,
of the operator $A$ given by
\beq
\Omega_{\vec k}&=& \left(\frac{2\pi} L n\right)^2+\left(\frac{2\pi}{\beta}
l\right)^2+\lambda_jR_H^{-2}+m^2+\xi R+\frac{\lambda} 2 \phi ^2\nn\\
&=:&\left(\frac{2\pi} L n\right)^2+\left(\frac{2\pi}{\beta}
l\right)^2+(r_j^2+\delta^2)R_H^{-2},\quad
j\in\nats _0, n,l\in\intgs\label{8}
\eeq
where we introduced the notations
$\de ^2=(M^2+\xi R)R_H^2+\frac 1 4$ and $M^2=m^2+\frac{\lambda} 2 \phi^2$.
Here $\lambda_j$ are the eigenvalues of the Laplacian $-\Delta_H$ of
the hyperbolic spatial part $H^2/\Gamma$ of the manifold ${\cal M}$,
normalized to $R_H=1$, where $R=-2R_H^{-2}$.

The contribution of this part of the spectrum of the one-loop
effective potential may be determined using the Selberg trace formula.
This will be done in the following.

First taking the limit
$L,\beta\to\infty$, we find
\beq
\lim_{L,\beta\to\infty}\frac{\zeta_A(s)}{L\beta}=\frac 1
{4\pi}R_H^{2(s-1)}F(s;\de )\label{9}
\eeq
where we introduced the function
\beq
F(s;\de )=\frac 1 {s-1} \sum_j\left[r_j^2+\de
^2\right]^{1-s}\label{10}
\eeq
The series is absolutely convergent for $\Re s >2$.
The necessary analytical continuation of equation (\ref{10}) to $s=0$
will be constructed using the Selbergs trace formula. Assuming
that $h(r)$ is analytic in the strip $|Im\,\,r|<1/2+\epsilon$,
$\epsilon>0$ and that $h(r) ={\cal O}((1+|r|^2)^{-1-\epsilon})$ in the
strip for $r\to\infty$,
it reads
\cite{selberg56}, \cite{venkov82}, \cite{hejhal76}
\beq
\sum_j h(r_j)&=&\frac{V({\cal F})}{4\pi}\iuu dr r\tanh(\pi r)h(r)\nn\\
& &+\sumh \sumkeu \frac{\ch l(P)}{2\sinh (kl(P))}\hat{h}
(kl(P))\label{11}\\
& &+\sume \sumkem \frac{\chq}{2n\sik}\iuu dr\frac{e^{-\frac{2\pi rk}
n}}{1+e^{-2\pi r}}h(r)\nn
\eeq
Here ${\cal F}$ is a fundamental domain and its measure can be
computed by the Gauss-Bonnet formula \cite{venkov82}, $V({\cal F})$ is
the volume of the fundamental domain and $\hat h$ is the Fourier
transform of $h$.
The summation $\{P\}_{\Gamma}$ ($\{Q\}_{\Gamma}$) is
taken over all primitive hyperbolic (elliptic) conjugacy
classes in $\Gamma$, $l(P)$ is the length of a closed geodesic
associated with the element $P$ of the conjugacy class,
$\chi$ is an arbitrary finite-dimensional representation of $\Gamma$
(character of $\Gamma$)
and finally $n=n(Q)$ is the order of the class with
representative $Q$. On the left-hand side the summation is over all
solutions $r_j$ of the equations $r_j^2=\lambda_j-1/4$.

Using equation (\ref{11}), the analytic continuation of equation
(\ref{10}) is found to be
\beq
F(s;\de )&=&\frac{V({\cal F})}
          {4\pi(s-1)(s-2)}\de ^{2(2-s)}-\frac{V({\cal F})}
          {2\pi (s-1)}\iou
          dr r(1-\tanh\pi r)\rd ^{1-s}\nn\\
& &+\frac{(2\de )^{\frac 3 2 -s}}{\sqrt{\pi}\Gamma(s)}\sumh\sumkeu
\frac{\ch l(P)}{2\sinh (kl(P))}(kl(P))^{s-\frac 3 2}K_{s-\frac 3 2}
(\de kl(P))\label{12}\\
& &+\sume \sumkem \frac{\chq}{2n\sik}\iuu dr \frac{e^{-\frac{2\pi rk}
n}}{1+e^{-2\pi r}}\rd ^{1-s}\nn
\eeq
The contribution to the effective potential associated with the
summation over hyperbolic classes, which we shall denote as $V_P$, can
be rewritten in terms of the logarithmic derivative of the Selberg
type zeta-function
\beq
\frac{Z'} Z (s)=\sumh \sumkeu \frac{\ch l(P)}{2\sinh (kl(P))}
e^{-(s-\frac 1 2 )kl(P)},\label{selberg}
\eeq
in the form \cite{bytsenkozerbini92,bytsenkovanzozerbini92a}
\beq
V_P=\frac 1 {\Gamma (s-1) \Gamma (2-s)}
\iou dy (y^2+2y\de )^{1-s} \frac{Z'} Z \left(y+\de +\frac 1 2
\right)\label{14}
\eeq
Now using equation (\ref{9}) and equation (\ref{6}) all the
information needed for the calculation of the one-loop quantum
correction is known, the only thing to do is to find the derivative of
equation (\ref{12}) at $s=0$. We will skip this intermediate step and
go directly to the effective potential. In order to give the final
result, it is useful to introduce some abbreviations. We will use
\beq
X&=&\iou dr r \rd \ln\erd (1-\tanh \pi r),\nn \\
H &=& \iou dy (y^2+2y\de)\frac{Z'} Z \left(y+\de +\frac 1
2\right),\nn\\
E &=&\iuu dr \frac{e^{-\frac{2\pi rk}
n}}{1+e^{-2\pi r}}\rd\ln\erd\nn\\
f(k)&=&\frac{\chq}{2n\siqk}\nn\\
h(k)&=&-\frac 1 4 +\frac 1 {2\siqk}\nn\\
g(k)&=&\frac{\chq}{n\sik}\nn
\eeq
In terms of these, the non-renormalized one-loop effective potential,
now including the for our calcutation necessary counterterms
\cite{utiyamawitt62}, \cite{toms82},
\cite{buchbinderodintsovshapiro89}, \cite{oconnorhu84}, is
\beq
V(\phi)&=&\Lambda +\de \Lambda +(\kappa +\de \kappa )R +
\frac 1 2 (\xi +
\de\xi )R\phi^2
+\frac 1 2 (a +\de a )R^2+\frac 1 {4!}(\lambda +\de \lambda )\phi
^4\nn\\
& &+\frac 1 2 (m^2+\de m^2)\phi^2+V_{hyp}(\phi) +V_{ell}(\phi)\label{13}
\eeq
where for clarity we introduce the hyperbolic (respectively
elliptic) contribution $V_{hyp}(\phi)$ (respectively $V_{ell}(\phi)$),
\beq
V_{hyp}(\phi)&=&-\frac{\hbar}{32\pi^2 R_H^4}\left\{\frac 1 2 \de
^4\left[\frac 3 2 +\ln\left(\frac{\mu^2
R_H^2}{\de^2}\right)\right]-2X+\frac {4\pi H} {V({\cal F})}
   \right.\nn\\
& &\left.\qquad\qquad+2\left[\frac 1 {24} \de^2 +\frac 7
{960}\right]\left[1+\ln\left(\frac{\mu^2 R_H^2}{\de
^2}\right)\right]\right\}\label{15}\\
V_{ell}(\phi)&=&-\frac{\hbar}{8\pi R_H^4 V({\cal F})}
\sume\sumkem\times\label{16}\\
& &\qquad\qquad\left\{-f(k)[\de
^2+h(k)]\left[1-\ln\left(\frac{\mu^2R_H^2}{\de^2}\right)\right]
       +g(k)E\right\}\nn
\eeq
In order to remove the dependence on the arbitrary  parameter $\mu$,
let us now continue with the renormalization procedure.
\section{Renormalization}
\setcounter{equation}{0}
The renormalization of the one-loop effective potential of a
self-interacting scalar field in curved spacetime is by now well known
(see for example \cite{oconnorhu84},
\cite{buchbinderodintsovshapiro89},
\cite{buchbinderodintsovshapiro92}), so our prescription will be
brief. We fix the counterterms by the renormalization condition
\beq
\xi &=& \frac{\abl ^3 V}{\abl R\abl \phi ^2}\left|_{R=0, \phi
=M}\right.\nn\\
a&=&\frac{\abl ^2 V}{\abl R^2}\left|_{R=0,\phi =M}\right.\label{17}\\
\lambda&=&\frac{\abl ^4 V}{\abl \phi^4}\left|_{R=0,\phi =M}\right.\nn
\eeq
For simplicity we restrict to the massless case $m=0$
(so no $\Lambda$ and $\kappa$ renormalization is necessary),
because our
main aim is to analyze the influence of the curvature and toplogy on
the phase transition. But in principle, nothing can prevent us from
adding a nonvanishing mass, the calculation would only be slightly
more difficult.

Using equation (\ref{13}), the counterterms are found to be
\beq
\de a&=&\frac{\hbar}{128 \pi^2}\left\{-4\left[\left(\xi -\frac 1 6
\right)^2+\frac 1 {180}\right]\lm\right.\nn\\
& &\qquad\qquad\left.+\frac {8\pi} {V({\cal F})}\sume\sumkem f(k)[\epsilon
+h(k)]\lm\right\}\nn\\
\de
\lambda&=&-\frac{\hbar\lambda^2}{32\pi^2}\left[8+3\lm\right]\label{18}\\
\de \xi
&=&-\frac{\hbar\lambda}{32\pi^2}\left[2+\lm\right]
    \left\{\xi -\frac 1 6 +\frac {2\pi} {V({\cal F})}
    \sume\sumkem f(k)\right\}\nn
\eeq
It is seen, that the given counterterms differ from the counterterms
derived in \cite{oconnorhu84} for a self-interacting scalar field
theory on a smooth manifold. This may be traced back to the co-compact
group $\Gamma$ under consideration,
due to which
the manifold $H^2/\Gamma$ is no longer a smooth one.

Introducing $\epsilon =\xi-\frac 1 8$,
after some calculation the final result for the
one-loop renormalized effective potential reads
\beq
V_r(\phi)&=&\frac 1 2 \xi R
\phi^2 +\frac 1 2 a R^2 +\frac{\lambda}{4!}\phi^4\nn\\
& &-\frac{\hbar}{128\pi^2}\times\nn\\
& &\left\{\frac{25}{12}\lambda^2 \phi^4 -\lambda \phi^2 R\left[\frac 9
8 -7\xi\right]+R^2\left[3\xi ^2 -\frac{11}{12}\xi
+\frac{79}{960}\right]\right.\nn\\
& &-2XR^2+\frac{4\pi HR^2}{V({\cal F})}\nn\\
& &-\left[\frac 1 2 \lambda^2\phi ^4+2\lambda \phi^2 R \left(\xi
-\frac 1 6 \right)+2R^2 \left(\left(\xi -\frac 1 6 \right)^2+\frac 1
{180}\right)\right]\times\nn\\
& &\qquad\qquad\left[\ln\left(\frac{\phi^2}{M^2}\right)+
\ln\left(1-\frac{\epsilon R}{\lambda \phi^2}\right)\right]\nn\\
& &+\frac {4\pi} {V({\cal F})}\sume \sumkem \times\label{19}\\
& &\left[3\lambda\phi^2 R f(k) -R^2 f(k) (\epsilon +h(k))+R^2
g(k)E\right.\nn\\
& &\left.\left.-f(k)\left(\lambda \phi^2 R -[\epsilon +
h(k)]R^2\right)\left(\ln\left(\frac{\phi^2}{M^2}\right)
+\ln\left(1-\frac{\epsilon R}{\lambda
\phi^2}\right)\right)\right]\right\}\nn
\eeq
For the subsequent analysis of the phase transition, let us consider several
limits. The small curvature case is very easily extracted from equation
(\ref{19}), we find
\beq
V_r(\phi)&=&\frac 1 2 \xi R
\phi ^2  +\frac 1 2 a R^2 +\frac{\lambda}{4!}\phi ^4\nn\\
& &-\frac{\hbar}{128\pi^2}\times\label{20}\\
& &\left\{\frac 1 2 \lambda^2 \phi^4
\left[\frac{25} 6-\ln\left(\frac{\phi^2}{M^2}\right)
\right]\right.\nn\\
& &+R\lambda
\phi^2\left[3-\ln\left(\frac{\phi}{M^2}\right)\right]\times\nn\\
& &\qquad\left[2\left(\xi -\frac 1 6\right)+\frac {4\pi}
      {V({\cal F})}\sume\sumkem
       f(k)\right]\nn\\
& &-R^2\ln\left(\frac{\phi^2}{M^2}\right)\times\nn\\
& &\left[2\left(\left(\xi -\frac 1 6\right)^2 +\frac 1 {180}\right)
    -\frac {4\pi} {V({\cal F})}
    \sume \sumkem f(k) (\epsilon + h(k))\right]\nn\\
& &\left.+{\cal O}(R^3)\right\}\nn
\eeq
Let us once more stress the importance of the presence of the elliptic
elements of $\Gamma$
leading to obvious contributions in (\ref{20}).
The small background field limit is
more difficult to obtain. We will
not state explicitly the constant part $\Lambda _{eff}$ of the
potential
(the so called cosmological constant),
but concentrate only on the quadratic contribution.
Introducing the functions
\beq
C(n)&=&\iou dr\;r^n(1-\tanh \pi r)\ln (\epsilon +r^2)\nn\\
F(n)&=&\iuu dr\;r^n \frac{e^{-\frac{2\pi rk} n}}{1+e^{-2\pi r}}\ln
(\epsilon +r^2)\nn\\
G(n)&=&\iou dy (y^2+2y\sqrt{\epsilon})^n\frac{Z'} Z
\left(y+\sqrt{\epsilon}+\frac 1 2\right)\nn
\eeq
the final expansion may be given in the form
\beq
V(\phi)&=&\Lambda_{eff} +\frac 1 2 \xi R \phi^2 \label{21}\\
& &-\frac{\hbar \lambda R}{128 \pi^2} \phi^2\times\nn\\
& &\left\{7\xi -\frac 9 8 -\frac 1 {12}\ln\epsilon
+2C(1)+\frac{4\pi G(0)}{V({\cal F})}\right.\nn\\
& &+2\left(\xi -\frac 1 6\right)\ln\left(-\frac{\lambda M^2}{\epsilon
R}\right)+\frac 2 {\epsilon}\left(\left(\xi -\frac 1 6 \right)^2-\frac
1 {576}\right)\nn\\
& &\left. +\frac {4\pi} {V({\cal F})}
     \sume\sumkem\left[f(k)\left(2+\ln\left(-\frac{\lambda
M^2} R \right)\right)-g(k)F(0)\right]\right\}\nn\\
& &+{\cal O}(\phi^4)\nn
\eeq
Armed with these results let us now continue with the
analysis of the phase transition in the
considered theory.
\section{Curvature induced phase transition}
\setcounter{equation}{0}
In this section we will try to analyze the curvature induced phase
transitions in our model, taking into account the topology effects.
Let us first
work in linear curvature approximation with the effective potential
(\ref{20}),
\beq
V_r(\phi)&=& \frac{\lambda}{4!} \phi^4-\frac 1 2 \xi |R| \phi^2
+\frac{\lambda^2 \phi^4}{256\pi
^2}\left[\ln\left(\frac{\phi^2}{M^2}\right)
-\frac{25}{6}\right]\nn\\
& &-\frac{\lambda |R| \phi^2}{64 \pi^2}
\left[\ln\left(\frac{\phi^2}{M^2}\right)-3\right]\left[\left(\xi
-\frac 1 6 \right)+T\right]\label{41}
\eeq
where $T=(2\pi /V({\cal F}))\sume\sumkem f(k)$. We took into account
that the curvature $R$ is negative, $R=- 2/ R_H^2$, hence $|R|=2/R_H^2$.

Let us discuss now the possibility of curvature induced phase
transition of first order. In this case the order parameter $<\phi>$
at some critical curvature $R_c$ is quickly changed. The general
theory of such phase transitions in linear curvature approximation has
been developed in \cite{buchbinderodintsov85},
\cite{buchbinderodintsovshapiro92}. Following these references, we
introduce the dimensionless variables $x=\phi^2/M^2$, $y=|R|/M^2$. In
these variables, the effective potential (\ref{41}) is given in the form
\beq
\frac{V}{M^4}&=& \frac{\lambda x^2}{4!} -\frac 1 2 \xi y x +
\frac{\lambda ^2 x^2}{256 \pi ^2}\left(\ln x -\frac{25} 6 \right)
\nn\\
& &-\frac{\lambda x y}{64 \pi^2} (\ln x -3)\left[\left(\xi -\frac 1
6\right)+T\right]\label{42}
\eeq
Note that the topological correction $T$ appears on the equal foot
with the parameter $\xi$ in the quantum correction to the effective
potential.

The standard conditions of the first-order phase transitions are
\beq
V(x_c,y_c)=0,\qquad \frac{\partial V}{\partial x}\left|_{x_c,y_c}=0,
\qquad \frac{\partial ^2 V}{\partial x^2}\left|_{x_c,y_c}>0
\right.\right.\label{43}
\eeq
Now, one can analyze the conditions (\ref{43}) for the potential
(\ref{42}) along the analysis given \cite{buchbinderodintsov85}.
The result of this analysis is the following. The first two conditions
of (\ref{43}) for the effective potential (\ref{42}) are fulfilled.
However, the leading term in $\partial ^2 V/\partial x^2|_{x_c,y_c}$
is cancelled,
\beq
\frac{\partial ^2 V}{\partial x^2}\left|_{x_c,y_c}=0+{\cal O}(\lambda
^2)\right.\label{44}
\eeq
Hence, the one-loop approximation is not enough to answer the question
on the possibility of gravitational phase transition in linear
curvature approximation. (The non-leading ${\cal O}(\lambda ^2)$-term
in (\ref{44}) which in our one-loop analysis is negative, can be
modified by two-loop corrections. Hence, the quite complicated two-loop
effective potential is necessary in order to answer the question
about the possibility of the first-oder phase transition.)

In order to estimate still the influence of the topology on the phase
transitions we will consider now the scalar electrodynamics. Of
course, we are not going to repeat the above analysis of the effective
potential calculation in that case because it is not so
straightforward and quite complicated from the technical viewpoint.
Instead we will use the general structure of the effective
potential in the one-loop approximation (that we already know from
(\ref{20})).

Then, the one-loop effective potential is given by (for simplicity we
work in Landau gauge),
\beq
V_r (\varphi) &=& \frac{\lambda}{4!} \varphi^4 +\frac 1 {(16 \pi) ^2}
\left(\frac {10} 9 \lambda ^2 +12 e^4\right) \left(\ln
\frac{\varphi^2}{M^2}-\frac {25} 6 \right)\varphi^4
-\frac 1 2 \xi |R| \varphi^2\nn\\
& & -\frac{|R|
\varphi^2}{(4\pi)^2}
\left[\ln\left(\frac{\varphi^2}{M^2}\right)-3\right]
\left\{\frac 1 3 \lambda \left[\left(\xi -\frac 1
6 \right) +T_1\right] +\frac 1 4 e^2 (1+T_2)\right\}\label{45}
\eeq
Here the first two terms represent the Coleman-Weinberg potential
\cite{colemanweinberg73}, $e^2$ is the electrical charge, $\varphi ^2
=\phi^{\dagger}\phi$, $T_1,T_2$ are topological corrections of a
similar nature as in (\ref{42}). We don't need their exact form
here, but $T_1$
should coincide with $T$ in (\ref{42}) up to some overall
coefficient.

One natural choice which is often used for estimations in flat space
\cite{colemanweinberg73} is $\lambda \sim e^4$. Then (\ref{45})
simplifies,but it still contains the topological
correction $T_2$. The result of the analysis
of equation (\ref{43})
shows that gravitational
phase transition is possible with the critical values
\beq
x_c \approx \exp (-1),\qquad y_c \approx -\frac{24 e^4}{\xi (16\pi
)^2}x_c,\qquad \xi<0 \label{46}
\eeq
Hence, the topological corrections do not influence the gravitational
phase transition in linear curvature approximation.

Now let us discuss again $\lambda \phi^4$-theory. The small
$\phi$-expansion
(\ref{21}) has the typical form
\beq
V_r(\phi) =\frac{const}{R_H^4} +R_H^{-2}\phi^2\lambda A +{\cal O}
(\phi^4)\label{47}
\eeq
where $A$ is easily obtained from (\ref{21}) and has a quite complicated
form. One can see \cite{allen83}, when $A$ changes the sign there is
a second-order phase transition in the theory. In this case
the critical curvature radius $_c\!R_H^2$ at which the phase
transition occurs depends
strongly on the topology. Explicitly we find
\beq
_c\!R_H^2=\frac{2}{\lambda M^2}\exp\left\{\frac{32\pi^2
B}{\hbar \lambda [\xi -1/6 +T]}\right\}\label{48}
\eeq
with
\beq
B&=&\xi -\frac{\hbar \lambda}{64\pi^2}\times\label{49}\\
  & &\qquad\left\{7\xi -\frac 9 8 -\frac 1 {12} \ln\ep +2C(1) +\frac
{4\pi}{V({\cal F})}G(0)\right.\nn\\
& & -2\left(\xi -\frac 1 6\right) \ln \ep +\frac 2 {\ep}
\left(\left(\xi -\frac 1 6 \right)^2-\frac 1 {576}\right)\nn\\
& &\left.
    +4T-\frac{4\pi}{V({\cal F})} \sume\sumkem f(k) g(k) F(0)\right\}\nn
\eeq
showing the relevance of the topology on the transition point
$_c\!R^2_H$.
\section{Conclusions}
\setcounter{equation}{0}
In this paper we considered a self-interacting scalar field coupled to
gravity on the spacetime of the form $\reals ^2 \times H^2/\Gamma$.
First we concentrated on the calculation of the one-loop effective
potential of the theory given in equation (\ref{19}). Using this
potential, it was possible to analyze the first order phase transition
in the linear curvature approximation and the second order phase
transition in the small background field limit. Whereas for the first
order phase transition the influence of the topology
in the one-loop approximation is not visible
(this remark is also true for the scalar electrodynamics on
$\reals ^2 \times H^2/\Gamma$),
we find that the critical curvature $_c\!R_H^2$ for the second order
phase transition strongly depends on the topology, (\ref{48}).

Note that as a by-product of our result regarding the renormalization
of the effective potential, one can easily get the well-known running
coupling constants
\beq
\lambda (t) &=& \frac{\lambda}{1-\frac{3\lambda t}{16\pi^2}}\nn\\
\xi (t) &=& \frac 1 6 +\left(\xi -\frac 1
6\right)\left(1-\frac{3\lambda t}{16\pi^2}\right)^{-\frac 1 3}\nn
\eeq
The interpretation of the scaling parameter $t$ as a parameter of the
scale transformation of the metric $g_{\mu\nu}\to e^{-2t}g_{\mu\nu}$
has been given in [38-43]
\nocite{nelsonpanangaden82,buchbinderodintsov83}
\nocite{buchbinderodintsov83a,buchbinderodintsov84,parkertoms84}
\nocite{parkertoms84a}
(for a general review see \cite{buchbinderodintsovshapiro92}).
As one can see, when $t\to -\infty$ (infrared region) the $\lambda
\phi^4$ theory is asymptotically free and asymptotically conformally
invariant.
\section{Acknowledgements}
We thank Guido Cognola, Luciano Vanzo and Sergio Zerbini for
helpful discussions. A.A.~Bytsenko and K.~Kirsten are grateful to the
Faculty of Science and the Department of Physics of the University of
Trento for kind hospitality.
S.D.~Odintsov thanks JSPS (Japan) for financial support.
\newpage

\end{document}